%% file: main.tex
\documentclass[conference]{IEEEtran}
\usepackage{amsmath,amssymb,amsfonts}
\usepackage{tikz}
\usetikzlibrary{arrows.meta}
\usepackage{booktabs}
\usepackage{graphicx} 
\usepackage{pgfplots}
\usepgfplotslibrary{groupplots}

\usetikzlibrary{positioning,calc}
\usepackage{subcaption}
\usepackage[american]{circuitikz}

\newcommand\blfootnote[1]{
  \begingroup
  \renewcommand\thefootnote{}
  \footnote{#1}
  \addtocounter{footnote}{-1}
  \endgroup
}

\title{Device-Circuit Co-Design of Variation-Resilient Read and Write Drivers for Antiferromagnetic Tunnel Junction (AFMTJ) Memories}

\author{
    \IEEEauthorblockN{Yousuf Choudhary and Tosiron Adegbija\\Department of Electrical and Computer Engineering\\The University of Arizona, USA\\
    Email: ychoudhary@arizona.edu, tosiron@arizona.edu}
}

\begin{document}

\maketitle

\begin{abstract}
Antiferromagnetic Tunnel Junctions (AFMTJs) offer picosecond switching and high integration density for in-memory computing, but their ultrafast dynamics and low tunnel magnetoresistance (TMR) make state-of-the-art MRAM interfaces unreliable. This work develops a device-circuit co-designed read/write interface optimized for AFMTJ behavior. Using a calibrated SPICE AFMTJ model as a baseline, we identify the limitations of conventional drivers and propose an asymmetric pulse driver (PD) for deterministic picosecond switching and a self-timed sense amplifier (STSA) with dynamic trip-point tuning for low-TMR sensing. Our experiments using SPICE and Monte Carlo evaluations demonstrate that the proposed circuits preserve AFMTJ latency and energy benefits while achieving robust read/write yield under realistic PVT and 3D integration parasitics, outperforming standard MRAM front-ends under the same conditions.
\end{abstract}

\section{Introduction}
Magnetic Tunnel Junctions (MTJs) are the cornerstone of commercial non-volatile memory, yet they face fundamental speed and efficiency bottlenecks dictated by ferromagnetic switching dynamics \cite{Bian21_PVTAware_STTMRAM_Sensing}. \textit{Antiferromagnetic Tunnel Junctions (AFMTJs)} overcome these limitations through inter-sublattice exchange coupling, enabling picosecond-scale switching, intrinsic robustness to external magnetic fields, and superior energy efficiency \cite{jungwirth2016antiferromagnetic, shao2024antiferromagnetic}. \blfootnote{This work was supported in part by NSF Grant 2425567.}

However, AFMTJs are incompatible with standard STT-RAM front-ends due to their unique resistance-area (RA) products and lower tunnel magnetoresistance (TMR) \cite{shao2024antiferromagnetic}. Furthermore, integration into dense or 3D-stacked arrays introduces thermal gradients and parasitic variations that severely erode the already narrow sensing margins of AFMTJs \cite{sabry2011towards}.

This work bridges AFMTJ device potential and array-level reliability through a co-designed read/write interface specialized to AFMTJ physics. We develop (i) \textbf{STSA+}, a tunable and thermally adaptive sense amplifier for low-TMR sensing; (ii) \textbf{PD\_EQ+}, a tier-aware precharge/equalization path for dense (e.g., 3D) arrays; and (iii) \textbf{WD\_WRITE}, an asymmetric pulse driver enabling fast, deterministic switching. By modeling realistic PVT variations and array parasitics, we show that these circuits achieve robust operation where standard MTJ-based interfaces fail.

\textbf{Our key contributions include:}

\begin{enumerate}
    \item \textbf{STSA+: A trip-point-tunable, thermally adaptive sense amplifier:} We propose a Strong-ARM-derived SA (STSA+) with programmable trip-point and temperature compensation that reliably resolves AFMTJ's low-TMR read margins and sustains sub-10$^{-6}$ BER across aggressive PVT and tier-induced thermal gradients.
    \item \textbf{PD\_EQ+: A tier-aware precharge/equalization driver:} We introduce an adaptive PD/EQ front-end that adjusts equalization time and drive strength based on AFMTJ RA statistics and tier temperature, improving bitline centering and expanding the disturbance-free sensing window by 2$\times$-4$\times$.
    \item \textbf{Variation-resistance write driver:} We propose an asymmetric pulse-shaping write driver (WD\_WRITE) that delivers picosecond-class AFMTJ switching with negligible ($<$1\%) energy overhead and maintains sub-10$^{-6}$ WER under voltage and pulse-width variation.
    \item \textbf{Robust evaluation under realistic conditions:} We integrate a calibrated dual-sublattice AFMTJ SPICE model with 3D-array parasitics and perform large-scale Monte Carlo sweeps. Our read/write path maintains sub-10$^{-6}$ BER/WER with wide tolerance to variation in $V_r$, SA delay, write bias, and pulse width, significantly outperforming state-of-the-art MRAM front-ends.
\end{enumerate}

\section{Background}
\label{sec:background}


\subsection{MTJ vs. AFMTJ: Device-Level Distinctions}
Conventional MTJs rely on a single ferromagnetic free layer that switches under spin-transfer torque (STT) or spin-orbit torque (SOT), with dynamics governed by a single magnetization vector $\mathbf{m}$~\cite{kent2015new}. In contrast, AFMTJs utilize two antiparallel sublattices coupled by strong exchange interactions ($J_\mathrm{AF}$), enabling THz-scale regime switching dynamics~\cite{jungwirth2016antiferromagnetic}.

Important distinctions arise from AFMTJ's dual-sublattice structure: its state is defined by the N\'eel order parameter $\mathbf{L} = \mathbf{m}_1 - \mathbf{m}_2$, where coupled vectors $\mathbf{m}_1$ and $\mathbf{m}_2$ evolve coherently  \cite{shao2024antiferromagnetic}. This coupling enables faster and more deterministic switching than ferromagnets, while the antiparallel configuration offers high thermal stability and immunity to magnetic disturbance. However, the resulting TMR is lower than in conventional MTJs, complicating reliable readout~\cite{shao2024antiferromagnetic}.


\subsection{AFMTJ Device Model}
We extend the standard UMN MTJ model \cite{kim2015technology} to capture the dual-sublattice AFMTJ physics based on Mn\textsubscript{3}SnN geometry. The dynamics of each sublattice ($i{=}1,2$) are governed by the modified Landau-Lifshitz-Gilbert (LLG) equation:
\begin{equation}
\frac{d\mathbf{m}_i}{dt}
= -\gamma\,\mathbf{m}_i \times \mathbf{H}_{\mathrm{eff},i}
+ \alpha\,\mathbf{m}_i \times \frac{d\mathbf{m}_i}{dt}
+ \boldsymbol{\tau}_{\mathrm{SOT},i}
+ \boldsymbol{\tau}_{\mathrm{ex},i},
\end{equation}
where $\boldsymbol{\tau}_{\mathrm{SOT}}$ represents the spin-orbit torque. The inter-sublattice exchange coupling is modeled as:
\begin{align}
\mathbf{H}_{\mathrm{ex},1} &= \frac{J_{\mathrm{AF}}}{M_s}\,\mathbf{m}_2,\quad
\boldsymbol{\tau}_{\mathrm{ex},1} = -T_{\mathrm{AF}}\,(\mathbf{m}_1 \times \mathbf{m}_2), \\
\mathbf{H}_{\mathrm{ex},2} &= \frac{J_{\mathrm{AF}}}{M_s}\,\mathbf{m}_1,\quad
\boldsymbol{\tau}_{\mathrm{ex},2} = -T_{\mathrm{AF}}\,(\mathbf{m}_2 \times \mathbf{m}_1),
\end{align}
with $T_{\mathrm{AF}} = \gamma J_{\mathrm{AF}} / M_s$. Our SPICE implementation captures the resistance trajectory $R(t)$ to validate switching probability and TMR distributions against experimental data~\cite{chen2024twist}.

\subsection{Baseline Sensing Architecture}
As a baseline, we employ a Strong-ARM-based sense amplifier performing \emph{single-threshold} detection against a reference voltage $v_\mathrm{ref}$ (Fig. \ref{fig:sa_pd}a). Since prior AFMTJ studies focus on device/material-level behavior without complete peripheral circuits, we benchmark against a state-of-the-art MTJ-based sensing interface \cite{bian2021investigation}. While standard for high-TMR MRAM, this topology degrades AFMTJ constraints: the reduced TMR signal window is easily overwhelmed by offset voltages and process variations, and fixed trip-points fail to track the significant resistance shifts caused by thermal gradients in 3D stacks, leading to high BER \cite{bian2021investigation}. These limitations motivate the adaptive sensing techniques proposed in Section \ref{sec:circuit}. 

\subsection{Write and Precharge Front-Ends}
Bitline initialization is handled by either a basic Precharge Driver (PD) or a PD with Equalization (PD\_EQ) to mitigate initial offsets. However, conventional drivers use fixed pulse widths that are oblivious to AFMTJ's specific switching latency or the local temperature. In a densely-integrated context (e.g., 3D implementations), this non-adaptive behavior risks either write failures (under-driving) or excessive energy consumption (over-driving) as device kinetics shift with tier temperature \cite{ bian2021investigation}. 

\subsection{3D‑Integration and Parasitic Modeling}

To simulate a realistic array environment, we model the bitline as a distributed RC ladder incorporating segmented interconnect resistance, TSV capacitance, and tier‑dependent thermal gradients \cite{chang2014thermal} ($\Delta T = 75^\circ$C between bottom logic and top memory tiers). Table \ref{tab:configs} summarizes the key array parameters.

\section{Proposed Circuit Enhancements} \label{sec:circuit}

To realize reliable operation in AFMTJ‑based arrays under PVT, thermal gradients, and device variation, we propose enhanced peripheral circuits that extend conventional architectures to adapt to AFMTJ‑specific behavior. A high-level comparison between prior MRAM peripheral circuits and our proposed AFMTJ-specific interfaces is shown in Fig.~\ref{fig:interfaces_comparison}, highlighting the key architectural differences that motivate the STSA+, PD\_EQ+, and WD\_WRITE circuits introduced next.

\input{interfaces_comparison}
\input{updated_sa_pd_combo}

\subsection{Adaptive Self-Timed Sense Amplifier (STSA+)}
To maintain readout margins across PVT corners, we augment the standard Strong‑ARM latch with dynamic tuning mechanisms. The proposed STSA+ architecture is shown in Fig.~\ref{fig:sa_pd}a.

\noindent\textbf{Programmable offset (VOFF):} We introduce a tunable offset bias that shifts the effective trip point, allowing compensation for static RA/TMR variations, temperature shift, or AFMTJ resistance spread.  

\noindent\textbf{Thermal $g_m$ compensation:} By modulating the tail transistor drive strength ($g_m$) based on the local tier temperature, we stabilize sensing delay and offset tolerance against thermal shifts.  

\noindent\textbf{Body-bias trip‑point modulation:} Drawing on prior work \cite{patel2021body}, we utilize dynamic body biasing for fine-grained threshold adjustment, mitigating mismatch without the power penalty of active current injection.

The goal of the STSA+ is to guarantee sufficient readout margin under worst-case variation, enabling reliable sensing with reduced bitline swing and lower read energy.

\subsection{PD\_EQ+: Thermal‑Aware Precharge and Equalization Driver}
Effective bitline initialization is critical for reading low-TMR devices. We propose an adaptive driver (PD\_EQ+), shown in Fig.~\ref{fig:sa_pd}b, that addresses this through \textbf{adaptive equalization} and \textbf{tier-aware drive strength}. The equalization pulse width is tuned adaptively based on statistical AFMTJ resistance data, ensuring the bitline differential is fully zeroed before sensing despite resistance spreads. To counter thermal gradients in 3D stacks, the equalization drive strength is modulated by tier position, ensuring uniform precharge timing. 

\subsection{Asymmetric Write Driver}
Our write driver (WD\_WRITE) is co-designed to match AFMTJ switching physics by incorporating \textit{programmable pulse shaping} to align pulse width and edge rates with the device's transient dynamics, thereby minimizing ringing and overshoot \cite{zhao2009high}. It provides \textit{asymmetric drive capability} to compensate for potentially different energy barriers of $P{\to}AP$ vs. $AP{\to}P$ transitions, allowing independent tuning for write-0 and write-1 strengths. The driver also includes \textit{thermal compensation}, adjusting its drive strength to maintain adequate torque at elevated temperatures without causing disturb errors in neighboring cells. 

\section{Simulation Setup}
\label{sec:simsetup}

We evaluate the proposed designs using HSPICE simulations with a 28 nm CMOS process co-simulated with our calibrated AFMTJ model. To ensure robustness against dense (3D) integration challenges, we sweep key PVT and parasitic corners: $V_\mathrm{DD}$ ($0.8\text{--}1.2$\,V), temperature ($300\text{--}475$\,K), bitline resistance ($R_\mathrm{BL}=100\text{--}300\,\Omega$), bitcell capacitance ($C_\mathrm{BIT}=15\text{--}30$\,fF), and TSV parasitics ($C_\mathrm{TSV}=10\text{--}20$\,fF). Read simulations sweep SA trip-points, bitline swing, and precharge timing across operating corners. Metrics include BER (validated to $<10^{-6}$ via a 3.15M-sample Monte Carlo), SA decision latency, read energy, and PVT margin. Write simulations use transient magnetization-dynamics modeling to characterize bias voltage (0.5--1.2 V), pulse width (0.1--1.5 ns), and temperature effects, extracting write latency, write energy, and disturb margin (using WER $<10^{-6}$ threshold).

Our architectural modeling assumes a dense 3D-stacked configuration with three tiers (3D-3T), comprising three vertically integrated tiers, each containing local wordlines and bank periphery. This setup explicitly accounts for practical constraints such as TSV keep-out zones, tier-dependent thermal constraints, and per-tier periphery placement. We model 32 banks of 8 Mb each (45 nm, 80 $F^2$ cells), including full array, driver/sense circuits, routing overheads, and intra-tier bitline/wordline lengths. Table~\ref{tab:configs} summarizes the key array parameters used to benchmark the area and energy overheads of the proposed drivers.

\begin{table}[t]
\centering
\caption{3D–3T AFMTJ IMC tile parameters (8\,Mb/bank, 32 total banks).}
\scriptsize
\begin{tabular}{@{}l l@{}}
\toprule
\textbf{Parameter} & \textbf{Value} \\
\midrule
Tiers $T$ & 3 \\
Total banks $B_{\text{tot}}$ & 32 \\
Total tile area (mm$^2$) & 15.94 \\
AFMTJ cell size & $80F^2$ (45\,nm) \\
Replication factor $m$ & 2--3 \\
Avg.\ TSV hops / decision & 1--3 \\
Bitline length (tile, cells) & $\approx 256$ \\
Wordline length (tile, cells) & 2048 \\
SA/PD OP table & SPICE-derived, shared across tiers \\
Thermal tiering & Upper tiers SA-limited under $T_{\text{max}}$ cap \\
\bottomrule
\label{tab:configs}
\end{tabular}
\vspace{-10pt}
\end{table}

\section{Results}
\input{write_latency_energy_combo}

\subsection{STSA vs. STSA+ Under PVT Variation}

We evaluate the sensing behavior for both the baseline STSA~\cite{Bian21_PVTAware_STTMRAM_Sensing} and our proposed STSA+ across supply and temperature corners. Table~\ref{tab:pvt_metrics} reports read latency (T\textsubscript{READ}) and read energy (E\textsubscript{READ}) for standard SS/TT/FF corners. STSA+ reduces read energy by \textbf{4--7$\times$} while maintaining identical latency, enabled by its offset-tunable trip-point and thermally adaptive tail bias. These features allow STSA+ to operate with reduced bitline swing without sacrificing offset tolerance or stability under variation.

\begin{table}[ht]
\centering
\caption{Read latency and energy under PVT corners (64×64 array)}
\label{tab:pvt_metrics}
\begin{tabular}{lcccc}
\toprule
\textbf{Scheme} & \textbf{Corner} & \textbf{VDD (V)} & \textbf{T\textsubscript{READ} (ns)} & \textbf{E\textsubscript{READ} (fJ)} \\
\midrule
STSA   & SS @ -40°C & 0.81 & 0.6 & 0.0505 \\
STSA   & TT @ 25°C  & 0.90 & 0.8 & 0.0733 \\
STSA   & FF @ 85°C  & 0.99 & 0.9 & 0.111  \\
\midrule
STSA+  & SS @ -40°C & 0.81 & 0.6 & 0.0123 \\
STSA+  & TT @ 25°C  & 0.90 & 0.8 & 0.0128 \\
STSA+  & FF @ 85°C  & 0.99 & 0.9 & 0.0140 \\
\bottomrule
\end{tabular}
\end{table}

\subsection{Precharge/EQ Behavior Under Tier Temperature Gradients}

We next analyze precharge and equalization under a realistic 3D thermal gradient (25°C top tier, 100°C bottom tier). With fixed equalization windows, the baseline PD\_EQ~\cite{sabry2011towards} fails to fully center the bitline in hot tiers, producing skewed inputs to the sense path. Our PD\_EQ+, on the other hand, adapts both equalization pulse width and drive strength based on estimated temperature and AFMTJ RA statistics, achieving substantially tighter centering. Under worst-case gradients, PD\_EQ+ expands the disturbance-free operating boundary by $2\times \text{ to }4\times$ relative to PD\_EQ.

\subsection{Write Performance, Thermal Robustness, and Overhead}
Fig. \ref{fig:combined_latency_energy} compares AFMTJ and MTJ write behavior across bias voltages. AFMTJs exhibit uniformly lower write latency due to exchange-enhanced dynamics. At 0.7\,V, AFMTJ achieves a write delay of 283.2\,ps vs. 1963\,ps for MTJ (6.9$\times$ faster). AFMTJs also deliver lower write energy (37.58\,fJ vs. 201.3\,fJ at 0.7 V). Our proposed write driver (\texttt{WD\_WRITE}) delivers a 0.7\,V pulse with 16\,ps rise/fall time and only 0.245\,fJ energy, showing that the overhead from driver circuitry is negligible---$<$1\% of the total write energy.


\input{mc_combined}

\begin{table}[t]
\centering
\caption{AFMTJ variation tolerance at 95\% confidence. Values show symmetric margins (\%) before BER/WER violations. Worst-case across P/AP and write directions.}
\label{tab:afmtj_tol}
\scriptsize
\begin{tabular}{@{}lcccc@{}}
\toprule
\textbf{Corner} & $\pm\%\ V_r$ (BER) & $\pm\%\ t_{\mathrm{SA}}$ (BER) & $\pm\%\ V_w$ (WER) & $\pm\%\ \tau_w$ (WER) \\
\midrule
25$^\circ$C & 11 & 37 & 8 & 17 \\
85$^\circ$C & 10 & 31 & 9 & 17 \\
\bottomrule
\end{tabular}
\vspace{-10pt}
\end{table}

\subsection{Monte Carlo Reliability Across Read/Write Paths}
We run Monte Carlo simulations on a 64$\times$64 AFMTJ tile (3.15M samples per operating point) to quantify statistical robustness (see Fig.~\ref{fig:mc_transients_pvt}). Across PVT corners, the nominal operating point satisfies both $\text{BER}_{95} \leq 10^{-6}$ and $\text{WER}_{95} \leq 10^{-6}$ at 25$^\circ$C and 85$^\circ$C. Table~\ref{tab:afmtj_tol} summarizes the 95\% variation margins: AFMTJs tolerate up to $\pm11\%$ read bias error, $\pm37\%$ SA delay deviation, $\pm8$–$9\%$ write voltage variation, and $\pm17\%$ write-pulse width variation. These margins confirm that the co-designed STSA+ and PD\_EQ+ maintain reliable detection and deterministic switching under realistic array-level variation.

\section{Conclusion}
In this work, we demonstrated that AFMTJ memories require deliberate circuit-device co-design to fully exploit their speed and efficiency advantages. By introducing the \textbf{STSA+} sense amplifier and \textbf{PD\_EQ+} precharge/equalize path, we address the low-TMR read margin and tier-dependent temperature imbalance that undermine conventional MRAM interfaces, enabling robust sensing across aggressive PVT and 3D-stack gradients. Together with \textbf{WD\_WRITE} driver, our interface achieves sub-10\textsuperscript{-6} BER/WER, preserves picosecond-scale AFMTJ switching, and maintains sub-pJ energy even under statistical variation. These results establish AFMTJ-based macros as practical, high-reliability, and energy-efficient components for future dense, latency-critical, and high-throughput VLSI systems.

\bibliographystyle{IEEEtran}
\bibliography{references}

\end{document}

%% file: interfaces_comparison.tex
\begin{figure}[t]
\centering
\resizebox{0.95\linewidth}{!}{
\begin{tikzpicture}[
    boxold/.style={draw, rounded corners, thick, minimum width=3.0cm, minimum height=1.0cm, align=center, fill=white},
    boxnew/.style={draw, rounded corners, thick, minimum width=3.2cm, minimum height=1.0cm, align=center, fill=blue!15},
    note/.style={font=\large, align=left},
    >=latex
]

\node[font=\bfseries] at (-2,4.3) {Prior Work (MTJ Interfaces)};
\node[font=\bfseries] at (4.5,4.3) {This Work (AFMTJ Interfaces)};

\node[boxold] (sa_old) at (-2,3.2) {Strong-ARM SA};
\node[note, right=0.2cm of sa_old] (sa_old_note) {Fixed threshold\\No dynamic ref};

\node[boxold] (pd_old) at (-2,1.8) {PD / PD-EQ};
\node[note, right=0.2cm of pd_old] (pd_old_note) {Fixed EQ window\\No thermal comp};

\node[boxold] (wd_old) at (-2,0.4) {Write Driver};
\node[note, right=0.2cm of wd_old] (wd_old_note) {Fixed pulse width\\No asymmetry};

\node[boxnew] (sa_new) at (4.7,3.2) {STSA+};
\node[note, right=0.2cm of sa_new] {Tunable VOFF\\Dynamic ref tracking\\$g_m$ thermal compensation};

\node[boxnew] (pd_new) at (4.7,1.8) {PD\_EQ+};
\node[note, right=0.2cm of pd_new] {Adaptive equalization\\Tier-aware drive strength};

\node[boxnew] (wd_new) at (4.7,0.4) {WD\_WRITE};
\node[note, right=0.2cm of wd_new] {Programmable pulse width\\Asymmetric write paths\\Thermal compensation};

\draw[->, thick] (sa_old.east) -- (sa_new.west);
\draw[->, thick] (pd_old.east) -- (pd_new.west);
\draw[->, thick] (wd_old.east) -- (wd_new.west);

\end{tikzpicture}
}
\caption{High-level comparison of conventional MRAM peripheral circuits and the AFMTJ-specific interfaces proposed in this work. The enhanced peripherals introduce tunability, thermal awareness, and adaptive timing to address the low-TMR signal window, as well as the variation-induced sensing and write challenges present in AFMTJ arrays.}
\label{fig:interfaces_comparison}
\end{figure}
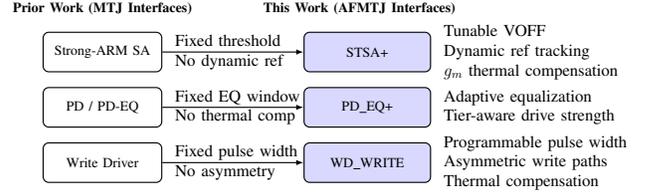

%% file: updated_sa_pd_combo.tex
\begin{figure}[t!]
\centering

\resizebox{0.7\columnwidth}{!}{\input{updated_stsa.tex}}
\vspace{-0.6em}
\centerline{\small (a) STSA+ sense interface}

\vspace{-3pt}

\resizebox{0.7\columnwidth}{!}{\input{updated_pd.tex}}
\vspace{-0.6em}
\centerline{\small (b) PD\_EQ+ precharge/equalization front-end}

\vspace{0.6em}
\caption{Peripheral circuits for AFMTJ sensing. (a) STSA+ augments a StrongARM latch with programmable offset injection, dynamic reference tracking, and temperature-aware $g_m$/tail-current biasing. (b) PD\_EQ+ CMOS switch-level front-end for initializing the bitline (BL) prior to evaluation.}
\label{fig:sa_pd}
\end{figure}

%% file: updated_STSA.tex
\centering
\begin{tikzpicture}[font=\large]

\begin{scope}[xshift=0cm, yshift=0cm,
  every node/.style={font=\Large},
  line/.style={thick},
  box/.style={draw, thick, rounded corners=2pt, fill=white},
  lab/.style={font=\large},
  slab/.style={font=\Large}
]

\coordinate (LT) at (6.2,5.8);
\coordinate (RT) at (8.2,5.8);
\coordinate (LM) at (6.2,3.9);
\coordinate (RM) at (8.2,3.9);
\coordinate (LB) at (6.2,2.1);
\coordinate (RB) at (8.2,2.1);
\coordinate (BOT) at (7.2,1.25);

\draw[line]
  (LT) -- (RT) -- (RM) -- (RB) -- (BOT) -- (LB) -- (LM) -- cycle;

\draw[line] (6.60,3.25) -- (7.80,2.65);
\draw[line] (7.80,3.25) -- (6.60,2.65);

\coordinate (TOPMID) at (7.2,6.15);
\draw[line] (LT) -- ++(0,0.35) coordinate (LPRE);
\draw[line] (RT) -- ++(0,0.35) coordinate (RPRE);
\draw[line] (LPRE) -- (RPRE);
\draw[line] (RPRE) -- (TOPMID);
\node[lab] at ($(TOPMID)+(0,0.20)$) {VDD};

\node[slab, above] at (LPRE) {$\overline{\text{pre}}$};
\node[slab, above] at (RPRE) {$\overline{\text{pre}}$};

\coordinate (oP) at (6.2,3.35);
\draw[line] (oP) -- ++(-0.4,0) node[left, lab] {$o^{+}$};
\coordinate (oM) at (8.2,3.35);
\draw[line] (oM) -- ++(0.40,0) node[right, lab] {$o^{-}$};

\coordinate (BLin) at (6.2,2.45);
\draw[line] (BLin) -- ++(-1.2,0) node[left, lab] {BL};

\coordinate (VRin) at (8.2,2.45);
\draw[line] (VRin) -- ++(0.55,0);

\node[lab, fill=white, inner sep=1.5pt] at ($(VRin)+(1.8,0.1)$) {$v_{\mathrm{ref,dyn}}$};

\draw[line] (BOT) -- ++(0,-0.70) coordinate (TAIL);
\coordinate (REN) at ($(BOT)+(0,-0.20)$);
\node[slab, right] at ($(REN)+(0.05,0.2)$) {ren};

\draw[line] (TAIL) -- ++(0,-0.25);
\draw[line] ($(TAIL)+(0,-0.25)$) -- ++(-0.18,-0.12);
\draw[line] ($(TAIL)+(0,-0.25)$) -- ++( 0.18,-0.12);
\draw[line] ($(TAIL)+(0,-0.37)$) -- ++(-0.13,-0.09);
\draw[line] ($(TAIL)+(0,-0.37)$) -- ++( 0.13,-0.09);

\node[box, minimum width=1.2cm, minimum height=0.85cm] (DAC) at (4.6,5.1) {};
\node[lab] at (DAC) {DAC};
\node[slab] at ($(DAC.north)+(0,0.25)$) {offset injection};

\node[lab] at (4.2,4.2) {$V_{\mathrm{OFF}}$};

\draw[line] (DAC.east) -- (6.2,5.1);

\node[box, minimum width=2.2cm, minimum height=1.0cm] (SHT) at (10.7,5.15) {};
\node[lab] at (SHT) {S/H + Track};
\node[slab] at ($(SHT.north)+(0,0.25)$) {ref tracking};

\draw[line] (SHT.east) -- ++(1.1,0) node[right, lab] {$v_{ref}$};

\draw[line] (SHT.west) -- ++(-0.35,0) coordinate (BUSh);
\draw[line] (BUSh) -- ++(0,-2.70);
\draw[line] ($(BUSh)+(0,-2.70)$) -- (8.75,2.45);

\node[box, minimum width=2.0cm, minimum height=1.0cm] (GMB) at (10.7,1.05) {};
\node[lab, align=center] at (GMB) {$g_m$-bias\\thermal};
\node[slab] at ($(GMB.south)+(0,1.25)$) {compensation};

\draw[line] (GMB.east) -- ++(1.1,0) node[right, lab] {TempCode};

\draw[line] (GMB.west) -- ++(-1.10,0) |- (REN);

\end{scope}

\end{tikzpicture}

\label{fig:stsa_pd_eq_plus}

%% file: updated_pd.tex
\centering
\resizebox{0.92\linewidth}{!}{%
\begin{tikzpicture}[font=\large]

\begin{scope}[
  every node/.style={font=\large},
  line/.style={thick},
  lab/.style={font=\small},
  slab/.style={font=\small},
  box/.style={draw, thick, rounded corners=2pt, fill=white}
]

\node at (6.5,6.5) {};

\coordinate (VDD) at (6.5,5.3);
\coordinate (BL)  at (6.5,2.7);
\coordinate (VREF)at (10.2,2.7);

\coordinate (Ptop) at (6.5,4.9);
\draw[line] (VDD) -- (Ptop);
\node[lab] at ($(VDD)+(0,0.25)$) {VDD};

\draw[line] (BL) -- ++(0,-1.7);
\node[lab, fill=white, inner sep=1.5pt, anchor=west] at (6.5,3) {BL};

\draw[line] (VREF) -- ++(0,0.0);
\node[lab, right] at (10.2,2.7) {$v_{ref}$};

\draw[line] (6.1,4.9) rectangle (6.9,4.2);
\node[slab] at (6.5,4.55) {P$_{PD}$};

\draw[line] (6.5,5.0) -- (6.5,5.3);
\draw[line] (6.5,4.2) -- (6.5,2.7);

\draw[line] (6.1,4.55) -- ++(-1.2,0);
\node[lab, left] at (4.9,4.55) {pc};

\node[lab, anchor=west, fill=white, inner sep=1pt]
  at (7.05,4.55) {\scriptsize PMOS precharge};

\draw[line] (8.1,3.1) rectangle (8.9,2.3);
\node[slab] at (8.5,2.7) {N$_{EQ}$};

\draw[line] (6.5,2.7) -- (8.1,2.7);

\draw[line] (8.9,2.7) -- (10.2,2.7);

\coordinate (NEQbot) at (8.5,2.3);
\draw[line] (NEQbot) -- ++(0,-0.85);
\node[lab] at ($(NEQbot)+(0,-1.05)$) {peq};

\node[lab, anchor=west, fill=white, inner sep=1pt]
  at (9.15,2.45) {\scriptsize NMOS eq};

\draw[line] (4.4,3.1) rectangle (5.1,2.3);
\node[slab] at (4.75,2.7) {P$_K$};

\draw[line] (4.75,3.1) -- ++(0,0.55) node[above, lab] {$v_{bias}$};
\draw[line] (5.1,2.7) -- (6.5,2.7);
\node[slab] at (4.75,2.0) {\scriptsize weak hold};

\node[lab, anchor=east, fill=white, inner sep=1pt]
  at (4.25,2.95) {\scriptsize weak PMOS keeper};

\end{scope}
\end{tikzpicture}
}
\label{fig:pd_eq_plus}

%% file: write_latency_energy_combo.tex
\begin{figure}[t!]
\centering
\vspace{-5pt}

\begin{tikzpicture}
\begin{axis}[
    hide axis,
    xmin=0, xmax=1,
    ymin=0, ymax=1,
    width=0.6\columnwidth,
    legend style={at={(0.5,1.1)}, anchor=south, legend columns=2, font=\scriptsize},
    legend cell align={left}
]
\addlegendimage{green!60!black, mark=triangle*}
\addlegendentry{AFMTJ (This Work)}
\addlegendimage{red, mark=square*}
\addlegendentry{MTJ (Baseline)}
\end{axis}
\end{tikzpicture}

\begin{subfigure}[t]{0.45\columnwidth}
\vspace{-100pt}
  \centering
  \resizebox{\linewidth}{!}{\input{write_latency}}
  \vspace{-15pt}
  \caption{Write latency}
  \label{fig:write_latency}
\end{subfigure}
\hfill
\begin{subfigure}[t]{0.45\columnwidth}
\vspace{-100pt}
  \centering
  \resizebox{\linewidth}{!}{\input{write_energy}}
  \vspace{-15pt}
  \caption{Write energy}
  \label{fig:write_energy}
\end{subfigure}

\vspace{-10pt}
\caption{Write performance comparison of AFMTJ vs. MTJ across input voltages.
Latency and energy include both the write and verify-read phases.}
\label{fig:combined_latency_energy}
 \vspace{-10pt}
\end{figure}

%% file: write_latency.tex
\centering

\begin{tikzpicture}
  \begin{axis}[
    scale only axis,
    xlabel={Input Voltage (V)},
    ylabel={Write Latency (ps)},
    grid=major,
    width=\linewidth,
    height=3cm,
    mark size=2pt,
    line width=1pt,
    xtick={0.5,0.6,0.7,0.8,0.9,1.0,1.1,1.2},
    yticklabel style={/pgf/number format/.cd, fixed, precision=0},
    ymin=100, ymax=4500,
  ]
    \addplot[
      color=green!60!black,
      mark=triangle*,
    ]
    coordinates {
      (0.5, 491.2)
      (0.6, 357.4)
      (0.7, 283.2)
      (0.8, 238.0)
      (0.9, 204.1)
      (1.0, 179.6)
      (1.1, 160.7)
      (1.2, 146.7)
    };

    \addplot[
      color=red,
      mark=square*,
    ]
    coordinates {
      (0.5, 4067)
      (0.6, 2517)
      (0.7, 1963)
      (0.8, 1631)
      (0.9, 1432)
      (1.0, 1332)
      (1.1, 1174)
      (1.2, 1089)
    };
  \end{axis}
\end{tikzpicture}
\label{fig:write_vs_vmtj}

%% file: write_energy.tex
\centering

\begin{tikzpicture}
\begin{axis}[
    scale only axis,
    width=\linewidth,
    height=3cm,
    xlabel={Input Voltage (V)},
    ylabel={Write Energy (fJ)},
    xmin=0.5, xmax=1.2,
    ymin=10, ymax=550,
    xtick={0.5,0.6,0.7,0.8,0.9,1.0,1.1,1.2},
    ytick={50,100,...,550},
    grid=both,
    grid style={line width=.1pt, draw=gray!30},
    major grid style={line width=.2pt,draw=gray!50},
    tick label style={font=\small},
    label style={font=\small},
    line width=1pt,
    mark size=2pt,
]

\addplot[
    color=green!60!black,
    mark=triangle*,
]
coordinates {
    (0.5, 30.02)
    (0.6, 32.35)
    (0.7, 37.58)
    (0.8, 44.91)
    (0.9, 53.88)
    (1.0, 55.98)
    (1.1, 62.11)
    (1.2, 67.26)
};

\addplot[
    color=red,
    mark=square*,
]
coordinates {
    (0.5, 67.0)
    (0.6, 133.4)
    (0.7, 201.3)
    (0.8, 290.8)
    (0.9, 427.3)
    (1.0, 507.3)
    (1.1, 493.2)
    (1.2, 500.7)
};
\end{axis}
\end{tikzpicture}
\label{fig:energy_vs_vmtj}

%% file: mc_combined.tex
\begin{figure}[t]
  \centering

  \begin{subfigure}[t]{0.45\columnwidth}
    \centering
    \input{read_mc_transient}
    \vspace{-15pt}
    \caption{Read path transient response}
    \label{fig:mc_read}
  \end{subfigure}
  \hfill
  \begin{subfigure}[t]{0.45\columnwidth}
    \centering
    \input{write_mc_transient}
    \vspace{-15pt}
    \caption{Write path transient response ($M_z$)}
    \label{fig:mc_write}
  \end{subfigure}

  \caption{
  Monte Carlo transient waveforms for a 64$\times$64 AFMTJ tile under PVT variation.
  (a) Read path: output voltage $v_\mathrm{out}$ from STSA+.
  (b) Write path: free-layer magnetization $M_z$ under PD\_EQ+ driving.
  }
  \label{fig:mc_transients_pvt}
  \vspace{-10pt}
\end{figure}
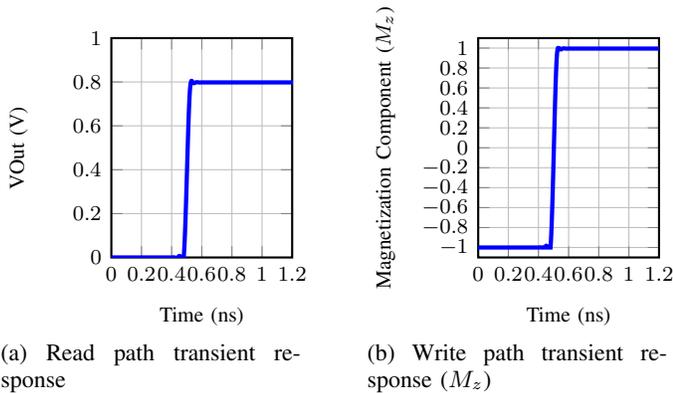

%% file: read_mc_transient.tex
\begin{tikzpicture}
\begin{axis}[
    width=\linewidth,
    height=4.5cm,
    xlabel={Time (ns)},
    ylabel={VOut (V)},
    xmin=0, xmax=1.2,
    ymin=0, ymax=1.0,
    xtick={0,0.2,...,1.2},
    ytick={0,0.2,...,1.0},
    grid=major,
    thick,
    legend style={at={(0.98,0.02)}, anchor=south east},
    every axis plot/.append style={ultra thick},
    tick label style={font=\footnotesize},
    label style={font=\footnotesize},
]
\addplot[
    color=blue,
    mark=none
] 
table[
    x=Time,
    y=Average_vout,
    col sep=comma
] {mc_transient_waveforms_read.csv};

\end{axis}

\end{tikzpicture}

%% file: write_mc_transient.tex
\begin{tikzpicture}
\begin{axis}[
    width=\linewidth,
    height=4.5cm,
    xlabel={Time (ns)},
    ylabel={Magnetization Component ($M_z$)},
    xmin=0, xmax=1.2,
    ymin=-1.1, ymax=1.1,
    grid=major,
    thick,
    legend style={at={(0.98,0.02)}, anchor=south east},
    every axis plot/.append style={ultra thick},
    xtick={0,0.2,...,1.2},
    ytick={-1,-0.8,...,1},
    tick label style={font=\footnotesize},
    label style={font=\footnotesize},
]
\addplot[
    color=blue,
    mark=none
] 
table[
    x=Time,
    y=Average_Mz,
    col sep=comma
] {mc_transient_waveforms_write.csv};

\end{axis}
\end{tikzpicture}